\documentclass[aps,pre,superscriptaddress,amsmath,amssymb,showpacs,twocolumn]{revtex4-1}
\usepackage{amssymb}
\usepackage{amsmath}
\usepackage{graphicx}
\usepackage{bm}
\usepackage{bbm}
\usepackage{color}
\usepackage{xcolor}
\usepackage{subfigure}

\newcommand{\z}[1]{\text{#1}}

\graphicspath{{./Figures/}}

\begin{document}

\title{Otto engine beyond its standard quantum limit}

\author{Bruno Leggio}
\affiliation{Laboratoire Charles Coulomb (L2C), UMR 5221 CNRS-Universit\'{e} de Montpellier, F- 34095 Montpellier, France}

\author{Mauro Antezza}
\affiliation{Laboratoire Charles Coulomb (L2C), UMR 5221 CNRS-Universit\'{e} de Montpellier, F- 34095 Montpellier, France}
\affiliation{Institut Universitaire de France, 1 rue Descartes, F-75231 Paris Cedex 05, France}

\newcommand{\ket}[1]{\displaystyle{|#1\rangle}}
\newcommand{\bra}[1]{\displaystyle{\langle #1|}}

\date{\today}

\begin{abstract}
We propose a quantum Otto cycle based on the properties of a two-level system in a realistic out-of-thermal-equilibrium electromagnetic field acting as its sole reservoir. This steady configuration is produced without the need of active control over the state of the environment, which is a non-coherent thermal radiation, sustained  only by external heat supplied to macroscopic objects. Remarkably, even for non-ideal finite-time transformations, it largely over-performs the standard ideal Otto cycle, and asymptotically achieves unit efficiency at finite power.
\end{abstract}


\maketitle

\section{Introduction}
Motivated by recent advancement in experimental techniques for the manipulation of single or few-body quantum systems \cite{Haenggi2009,Li2012,Haroche2013}, a thermodynamic description of microscale and nanoscale phenomena has been attracting a huge deal of attention \cite{GemmerBook}. Among its many different topics, one can notably list the quantum formulation of the laws of thermodynamics \cite{Brandao2008,Levy2012,Horodecki2013}, the physics of strongly non-equilibrium quantum dynamics \cite{Esposito2009,Leggio2013a,Leggio2013b}, the characterization of quantum thermal machines \cite{Linden2010,Venturelli2013,Leggio2015,Leggio2015b}, and the study of energy transport phenomena \cite{Bermudez2010,Leggio2015c,Wang2015}. All these research lines imply the descritpion of the interaction of quantum systems with large, usually classical environments. In particular, the interaction of quantum emitters with electromagnetic radiation has been largely studied both in equilibrium and non-equilibrium thermodynamic contexts: out-of-thermal equilibrium electromagnetic fields have been, for instance, shown to provide an ideal playground to induce and exploit stationary quantum properties in a many-emitters system.

One of the most promising outcome of quantum thermodynamics is the characterization of quantum-scale heat engines. These are quantum systems, referred to as working fluid, undergoing well-established cycles during which they interact with classical reservoirs and exchange work with an external device. In particular, the so-called Otto cycle is one of the main thermodynamic cycles, both in classical \cite{CallenBook} and quantum contexts \cite{GemmerBook}. Thanks to its theoretical simplicity, it allows to explore profound physical ideas, while still representing nowadays one of the most employed cycles, notably at the core of the functioning of many four-stroke engines. In quantum contexts, alongside the Carnot cycle, it has been a milestone of the development of a quantum thermodynamics formalism \cite{GemmerBook,Alicki1979,Kieu2004,Quan2007,Klimovsky2013}. Furthermore, many micro- and nanoscopic realizations of thermodynamic cycles have recently been proposed and achieved \cite{Scully2003,Abah2012,Hugel2002, Steeneken2011,Blickle2011,Bergenfeldt2014}.

In this paper, we employ non-equilibrium electromagnetic radiation to enhance the performances of the quantum Otto cycle (QOC) of a two-level light emitter. We show that, thanks to the realistic and non-trivial structure of such non-equilibrium reservoir for the quantum working fluid, both cycle efficiency and power output can largely overcome their standard equilibrium values. This work is structured as follows: in Section \ref{sqoc} we briefly review the definition and the physical properties of a standard equilibrium QOC for a two-level system. Section \ref{ote} is devoted to the description of the interaction of quantum emitters with a particular and realistic out-of-thermal equilibrium (OTE) electromagnetic field produced by a macroscopic object embedded in a thermal blackbody radiation; this will be employed in Section \ref{oteqoc} to give the main result of this paper, namely, a non-equilibrium quantum Otto cycle with remarkably high performances. Finally, remarks are given and conclusions are drawn in Section \ref{conc}.

\section{Standard quantum Otto cycle}\label{sqoc}
As any standard thermodynamic cycle, the Otto cycle happens between two temperatures imposed by ideal thermal reservoirs. Classically it consists of four stages: two isochoric processes during which the working substance exchanges heat with one of the two thermal reservoirs, and two adiabatic processes through which work is exchanged with the external world.
\begin{figure}[t!]
\begin{center}
\includegraphics[width=220pt]{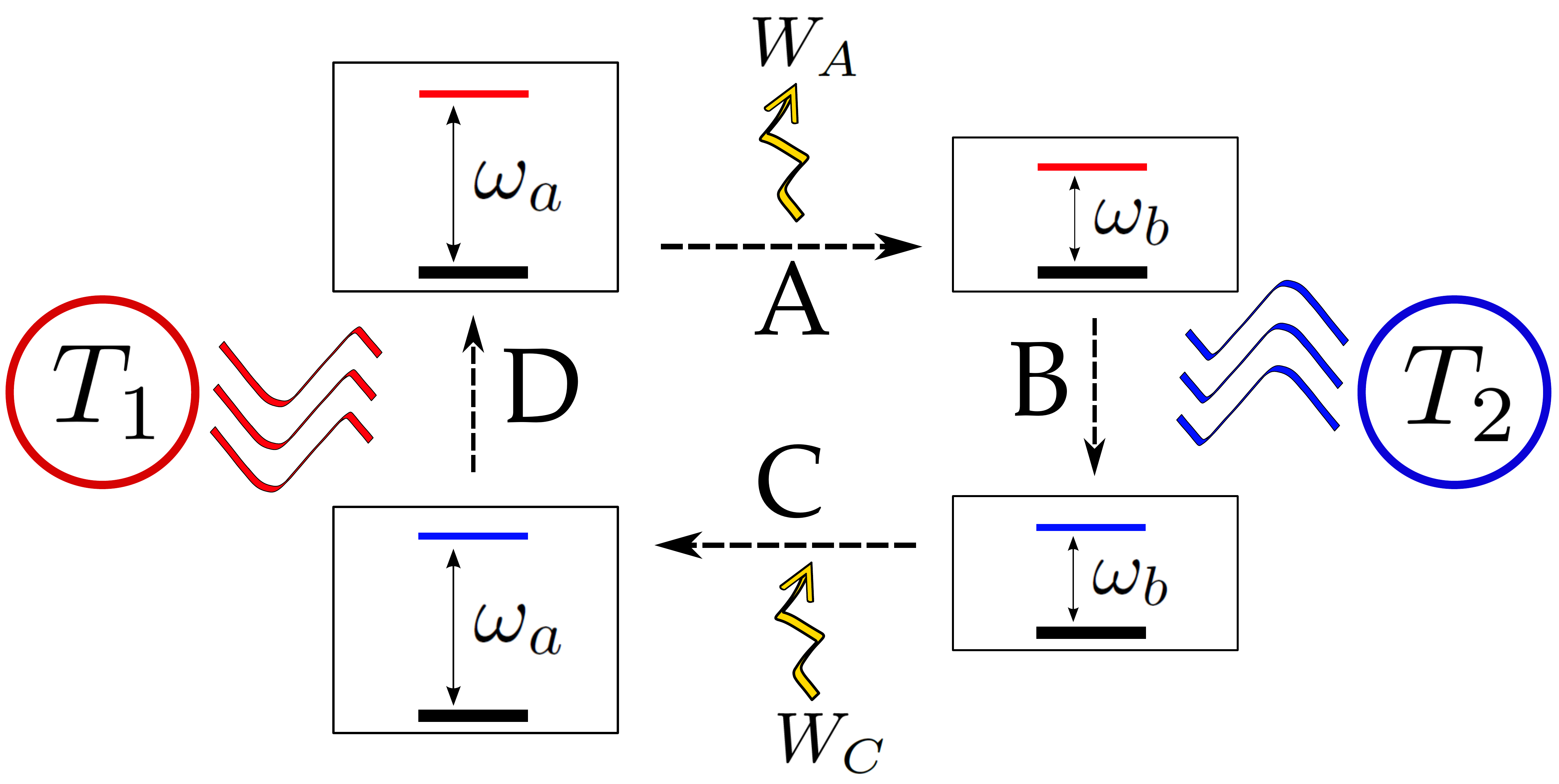}
\end{center}
\caption{Standard quantum Otto cycle. During adiabatic stages A and C the two-level working fluid exchanges work with the external world, while during stages B and D the TLS is put in contact with reservoirs at, respectively, $T_2$ and $T_1>T_2$.}
\label{Figure1S}
\end{figure}
Its quantum version for a quantum two-level system (TLS) as working fluid consists of four stages between two different temperatures $T_1>T_2$ \cite{Kieu2004,Wang2013} as schematically depicted in Fig.~\ref{Figure1S}. The standard quantum Otto cycle (s-QOC) is realized by directly putting the working fluid in contact with the thermal reservoirs in the equivalent of isochoric stages.
The internal energy $U$ of the TLS depends on two parameters only: its frequency $\omega$ (such that $\hbar \omega$ is the energy separation of its two levels), and the excited state population $p_e$. In particular, $U=\hbar \omega p_e$. Heat flowing into/out of the TLS will change $U$ by affecting $p_e$, whereas work contributions will change the value of $\omega$.

The working fluid is initially prepared, at frequency $\omega_a$, in a thermal state at temperature $T_1$ with excited state population $p_e(\omega_a,T_1)$, having introduced the excited state population of a two-level system of frequency $\omega$ and in equilibrium at temperature $T$ as $p_e(\omega,T)=\big(1+\exp[\hbar \omega/(k_B T)]\big)^{-1}$. The TLS then undergoes four transformations:
\begin{itemize}
\item A: ``expansion" $\omega_a\rightarrow\omega_b<\omega_a$. Since the energy of the TLS decreases, work is done by the fluid on the external world. Adiabaticity is given by the fact that $p_e=p_e(\omega_a,T_1)$ is constant;

\item B: thermalization of the system at frequency $\omega_b$ with the reservoir at low temperature $T_2$. No work is done by or on the system, which releases heat into the reservoir, changing its population to $p_e(\omega_b,T_2)$;

\item C: ``compression" $\omega_b\rightarrow\omega_a$. The energy of the TLS is now increased, such that work is exerted on the working fluid; as in A, the adiabatic assumption means that $p_e=p_e(\omega_b,T_2)$ stays constant;

\item D: thermalization of the system at frequency $\omega_a$ with a reservoir at high temperature $T_1$, such that the initial cycle condition is restored. No work is done by or on the system, which absorbs heat from the reservoir until $p_e=p_e(\omega_a,T_1)$.
\end{itemize}
The adiabaticity of stages A and C can be achieved by changing the frequency over a time interval much shorter than the one needed for the working fluid to interact with a thermal bath. In what follows, when thinking about the standard quantum description of the cycle, we will always have in mind the standard ideal (i.e., infinite frequency-tuning speed) quantum Otto cycle, referred to as si-QOC or simply QOC. In this configuration, the efficiency and the power delivered depend only on fundamental quantities, independently on the practical realization of the cycle \cite{Kieu2004,Wang2013}.

At the end of a si-QOC the net work made \textit{by the working fluid} (wf) \textit{on the external world} is given by the internal energy change during stages A and C:
\begin{equation}\label{wwf}
W_{\mathrm{wf}}=\hbar(\omega_b-\omega_a)p_e(\omega_a,T_1)+\hbar(\omega_a-\omega_b)p_e(\omega_b,T_2).
\end{equation}
On the other hand, the heat \textit{absorbed by the fluid} (stage D) reads
\begin{equation}
Q_{\mathrm{abs}}=\hbar\omega_a\Big(p_e(\omega_a,T_1)-p_e(\omega_b,T_2)\Big).
\end{equation}
Note that not just any value $\omega_b$ can be chosen. Indeed, for the cycle to be thermodynamically convenient one must require that $W_{\mathrm{wf}}<0$ (i.e., one is extracting net work from the system). This requirement leads to the so-called positive-work condition PWC which, directly from Eq.~\eqref{wwf}, reads $\omega_b/\omega_a\geq T_2/T_1$; moreover, the efficiency of work extraction defined as $\eta=-W_{\mathrm{wf}}/Q_{\mathrm{abs}}$ is readily evaluated as
\begin{equation}\label{eta}
\eta=1-\frac{\omega_b}{\omega_a}\leq 1-\frac{T_2}{T_1}=\eta_{\mathrm{C}},
\end{equation}
$\eta_{\mathrm{C}}$ being the Carnot efficiency between the same two temperatures $T_1$ and $T_2$. As such, the natural requirement that work extraction vanishes at the Carnot limit, i.e., $\eta=\eta_{\mathrm{C}}\Rightarrow W_{\mathrm{wf}}=0$, is obeyed, as one immediately verifies by using the condition $\omega_b=\omega_a T_1/T_2$ in Eq.~\eqref{wwf}.

Recently, however, it has been shown that the introduction of non-equilibrium features in the two reservoirs the working fluid interacts with in stages B and D can allow to go beyond these fundamental bounds \cite{Rossnagel2014,Abah2014,Alicki2015}. In particular, squeezing or, in general, coherence into electromagnetic reservoirs has been shown to provide higher cycle performances. Coherence requires however a detailed and steady control over the state of the baths, which can be cumbersome and usually implies the need of external work to be supplied. One would thus like to have an equivalent enhancement of cycle performances without the need of active control over the state of the reservoirs and, possibly, without the need of any work supply. In this work we propose a scheme to achieve this idea, by exploiting the out-of-thermal-equilibrium (OTE) properties of a realistic electromagnetic field produced by a body at a fixed temperature embedded in a blackbody radiation not in thermal equilibrium with it \cite{Messina2011,Messina2011b,Bellomo2013,Bellomo2013b,Bellomo2015}.

\section{Out-of-thermal-equilibrium field and its interaction with quantum emitters}\label{ote}
\begin{figure}[t!]
\begin{center}
\includegraphics[width=220pt]{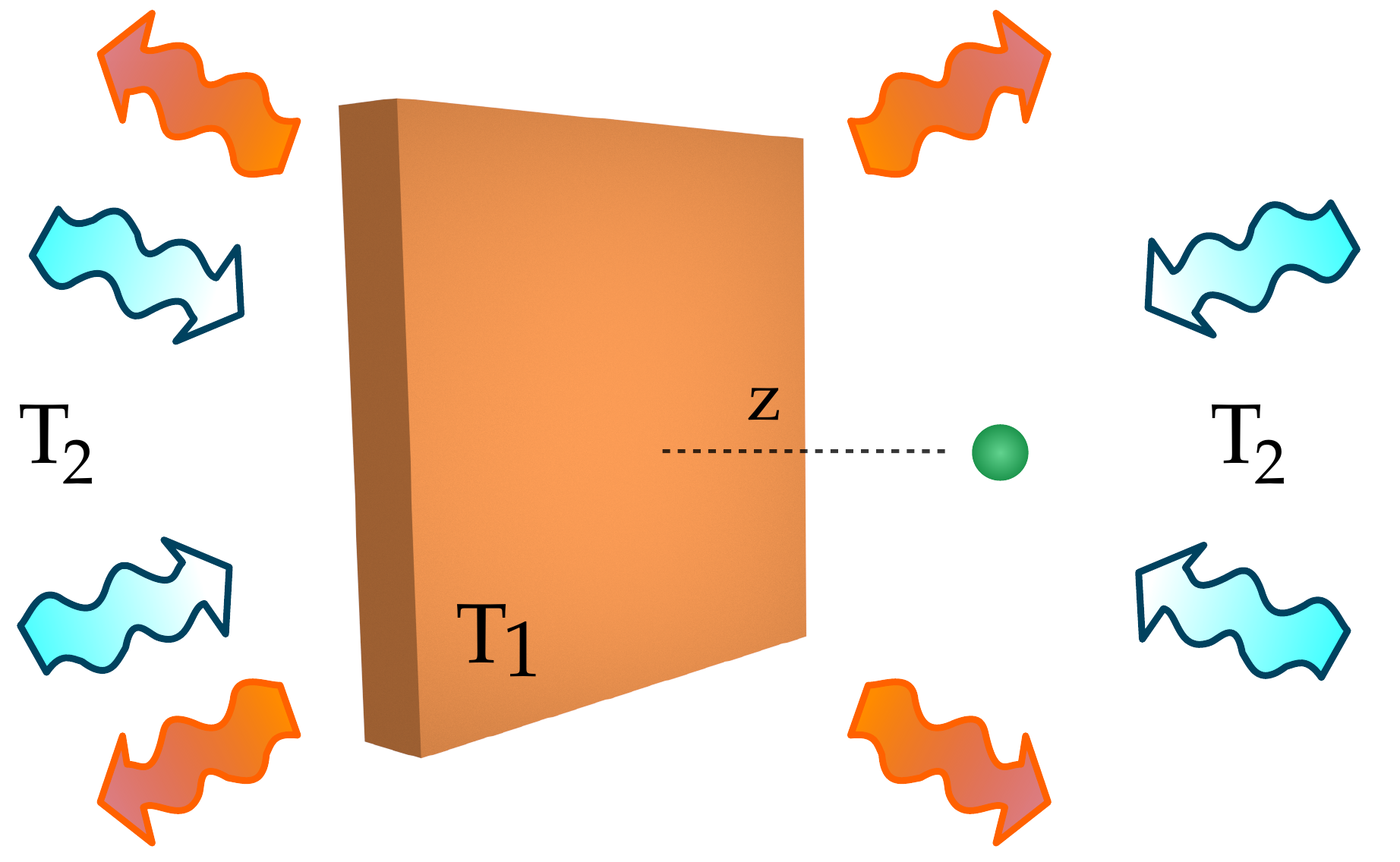}
\end{center}
\caption{Out-of-thermal equilibrium configuration. A slab of dielectric material is kept at a fixed temperature by means of a thermal reservoir at $T_1$, and embedded in a blackbody field at $T_2<T_1$. Quantum emitters placed at a distance $z$ from the slab surface interact with a non-trivial steady electromagnetic field.}
\label{OTEconfig}
\end{figure}

Let us then assume to have at disposal the same two thermal reservoirs at $T_1$ and $T_2<T_1$. Instead of directly coupling them to the working fluid, we suggest to employ them to produce an out-of-thermal equilibrium (OTE) electromagnetic field, whose features can be exploited to enhance the cycle performances.
Imagine thus to connect the reservoir at $T_1$ to a macroscopic object of some kind, for instance a slab of dielectric material of finite thickness $\delta$, and to embedded it in a thermal blackbody radiation at $T_2$, as depicted in Fig.~\ref{OTEconfig}. This configuration generates in the whole space around the slab a steady OTE field, whose properties depend on the dielectric and geometric properties of the slab through its reflection and transmission matrices. As such, the characterization of such a field is fully realistic when one employs real dielectric functions for the particular material of the slab.

When quantum emitters are placed in this field, they couple with it. In the dipolar approximation limit, this coupling has the form $H_I~=~-\sum_{i}\mathbf{d}_i \cdot \mathbf{E}(\mathbf{R}_i)$, where $i$ runs over all possible transitions of the quantum emitters and, in absence of permanent atomic dipoles, $\mathbf{d}_i$ is the field-induced dipole moment of the $i$-th transition, belonging to a quantum emitter located at $\mathbf{R}_i$.

In the weak coupling limit \cite{BreuerBook}, the dynamics of the sole atomic part can be described by a Markovian master equation \cite{Bellomo2013}. Be $\sigma_i^{-(+)}$ the lowering (raising) operator of transition $i$: the emitters master equation reads
\begin{equation}\label{MEtot}
\frac{d\rho}{d t}=-\frac{i}{\hbar}\big[H_\text{eff},\rho\big]+D_\text{loc}(\rho)+D_{\mathrm{nl}}(\rho),
\end{equation}
where $H_\text{eff}=\sum_i\omega_i\sigma_i^+\sigma_i^-+\sum_{i,j}^{\text{res}}\Lambda_{ij}\sigma_i^-\sigma_j^+$ is the effective emitters Hamiltonian, in which the free part is modified by a field-induced dipole-dipole coupling of strength $\Lambda_{ij}$ between two \textit{resonant} transitions $i$ and $j$. The sum $\sum_{i,j}^{\z{res}}$ runs over any possible pair of resonant transitions $i,j$.

The dissipative effects induced by the field are described by the dissipators $D_\text{loc}$ and $D_\text{nl}$, each given in terms of $\sigma_i^{\pm}$ as
\begin{eqnarray}
D_\z{loc}&=&\sum_i\left(\gamma_i^+L(\sigma_i^-)+\gamma_i^-L(\sigma_i^+)\right),\\
D_\z{nl}&=&\sum_{i,j}^\z{res}\left(\gamma_{ij}^+R(\sigma_i^-,\sigma_j^-)+\gamma_{ij}^-R(\sigma_i^+,\sigma_j^+)\right)\label{dnl},
\end{eqnarray}
having introduced the non-diagonal and diagonal lindblad dissipators as, respectively, $R(K_1,K_2)=K_1\rho K_2^{\dag}-1/2\big\{K_2^{\dag}K_1,\rho\big\}$ and $L(K)=R(K,K)$. All the rates $\gamma_i^{\pm}$, $\gamma_{ij}^{\pm}$ and $\Lambda_{ij}$ are directly obtained from the self-correlation functions of the electromagnetic field \cite{Bellomo2013} and depend on each transition frequency, dipole magnitude and orientation, on the geometric and dielectric properties of the slab and on the atom-atom and atom-slab distances.

The self-correlation functions of components $l\in\{x,y,z\}$ and $m\in\{x,y,z\}$ of the field at, respectively, point $R_i$ and $R_j$ in space are defined as
\begin{eqnarray}
c^{ij}_{lm}(\omega)&=&\frac{1}{\hbar^2}\langle E_l(R_i,\omega)E^{\dag}_m(R_j,\omega)\rangle,\label{corrfun1}\\
c^{ij}_{lm}(-\omega)&=&\frac{1}{\hbar^2}\langle E_l^{\dag}(R_i,\omega)E_m(R_j,\omega)\rangle.\label{corrfun2}
\end{eqnarray}
In what follows, it is more convenient to separate in $R_i$ the position $\mathbf{r}_i$ in the $x-y$ plane (parallel to the slab surface) from the $z_i$ position ($z=0$ being the coordinate of the slab surface), thus writing $R_i=\{\mathbf{r}_i,z_i\}$.
Functions \eqref{corrfun1}-\eqref{corrfun2} can be given an expression in terms of the slab and blackbody temperatures $T_1$ and $T_2$, and of the transmission and reflection scattering operators of the slab, which in turn depend on the thickness and dielectric permittivity of the slab material \cite{Messina2011,Bellomo2013}. Introducing the average photon number at frequency $\omega$ and temperature $T$ as  $n(\omega,T)=\Big[\exp{\big(\hbar \omega/k_B T}\big)-1\Big]^{-1}$, their explicit expressions read
\begin{eqnarray}
&\langle E_l(R_i,\omega)E^{\dag}_m(R_j,\omega)\rangle=\frac{\hbar \omega^3}{3\pi\varepsilon_0c^3}\Big\{\big[1+n(\omega,T_1)\big]\alpha_1^{i,j}(\omega)\big|_{lm}\nonumber \\
&+\big[1+n(\omega,T_2)\big]\alpha_2^{i,j}(\omega)\big|_{lm}\Big\},\label{corr1}\\
&\langle E_l^{\dag}(R_i,\omega)E_m(R_j,\omega)\rangle=\frac{\hbar \omega^3}{3\pi\varepsilon_0c^3}\Big\{n(\omega,T_1)\alpha_1^{i,j}(\omega)\big|_{lm}\nonumber \\
&+n(\omega,T_2)\alpha_2^{i,j}(\omega)\big|_{lm}\Big\},\label{corr2}
\end{eqnarray}
where the $3\times3$ matrices $\alpha_{1,2}^{i,j}(\omega)$ are given by
\begin{equation}\label{alpha1}
\begin{split}
\alpha_1^{i,j}(\omega)=&\frac{3\pi c}{2\omega}\sum_{p,p^{'}}\int\frac{d^2\mathbf{k}}{(2\pi)^2}\frac{d^2\mathbf{k'}}{(2\pi)^2}e^{i(\mathbf{k}\cdot \mathbf{r}_i-\mathbf{k'}\cdot \mathbf{r}_j)}\\
\times&\langle p,\mathbf{k}|\bigg\{e^{i(k_zz_i-k_z'^*z_j)}X_{p,p'}^{++}(\mathbf{k},\mathbf{k}',\omega)\\
\times&\Big(\mathcal{P}^{\text{pw}}_{-1}-\mathcal{R}\mathcal{P}^{\text{pw}}_{-1}\mathcal{R}^{\dag}-\mathcal{P}^{\text{ew}}_{-1}\mathcal{R}^{\dag}-\mathcal{T}\mathcal{P}^{\text{pw}}_{-1}\mathcal{T}^{\dag}\\
+&\mathcal{R}\mathcal{P}^{\text{ew}}_{-1}\Big)\bigg\}|p',\mathbf{k}'\rangle,
\end{split}
\end{equation}
\begin{equation}\label{alpha2}
\begin{split}
\alpha_2^{i,j}(\omega)=&\frac{3\pi c}{2\omega}\sum_{p,p^{'}}\int\frac{d^2\mathbf{k}}{(2\pi)^2}\frac{d^2\mathbf{k'}}{(2\pi)^2}e^{i(\mathbf{k}\cdot \mathbf{r}_i-\mathbf{k'}\cdot \mathbf{r}_j)}\\
\times&\langle p,\mathbf{k}|\bigg\{e^{i(k_zz_i-k_z'^*z_j)}X_{p,p'}^{++}(\mathbf{k},\mathbf{k}',\omega)\\
\times&\Big(\mathcal{T}\mathcal{P}^{\text{pw}}_{-1}\mathcal{T}^{\dag}+\mathcal{R}\mathcal{P}^{\text{pw}}_{-1}\mathcal{R}^{\dag}\Big)\\
+&e^{i(k_zz_i+k_z'^*z_j)}X_{p,p'}^{+-}(\mathbf{k},\mathbf{k}',\omega)\mathcal{R}\mathcal{P}_{-1}^{\text{pw}}\\
+&e^{-i(k_zz_i+k_z'^*z_j)}X_{p,p'}^{-+}(\mathbf{k},\mathbf{k}',\omega)\mathcal{P}_{-1}^{\text{pw}}\mathcal{R}^{\dag}\\
+&e^{-i(k_zz_i-k_z'^*z_j)}X_{p,p'}^{--}(\mathbf{k},\mathbf{k}',\omega)\mathcal{P}_{-1}^{\text{pw}}\bigg\}|p',\mathbf{k}'\rangle,
\end{split}
\end{equation}
being $k_z=\sqrt{\frac{\omega^2}{c^2}-\mathbf{k}^2}$, and where the operator $\mathcal{P}_{-1}^{\text{pw}(\text{ew})}$ is the projector on the propagative (evanescent) sector divided by $k_z$. We have introduced the $3\times3$ matrices $X_{p,p'}^{\mu\nu}(\mathbf{k},\mathbf{k}',\omega)\big|_{lm}=\hat{\mathbf{\epsilon}}_p^{\mu}(\mathbf{k},\omega)\big|_l\times \hat{\mathbf{\epsilon}}_{p'}^{\nu}(\mathbf{k}',\omega)\big|_m$, $\hat{\mathbf{\epsilon}}_p^{\mu}(\mathbf{k},\omega)$ being the polarization unit vector of the electromagnetic field, corresponding to polarization $p\in\{\text{TE},\text{TM}\}$ and z-component of the propagation direction $\mu\in[+,-]$ \cite{Messina2011}. The operators $\mathcal{R}$ and $\mathcal{T}$ describe, respectively, reflection and transmission of electromagnetic radiation by the slab and, as such, depend on the slab dielectric permittivity $\varepsilon(\omega)$ and slab thickness $\delta$ as
\begin{eqnarray}
\langle p,\mathbf{k}|\mathcal{R}|p',\mathbf{k}'\rangle&=&(2\pi)^2\delta(\mathbf{k}-\mathbf{k}')\delta_{pp'}\rho_p(\mathbf{k},\omega),\\
\langle p,\mathbf{k}|\mathcal{T}|p',\mathbf{k}'\rangle&=&(2\pi)^2\delta(\mathbf{k}-\mathbf{k}')\delta_{pp'}\tau_p(\mathbf{k},\omega),
\end{eqnarray}
with 
\begin{eqnarray}
\rho_p(\mathbf{k},\omega)&=&r_p(\mathbf{k},\omega)\frac{1-e^{2ik_{zm}\delta}}{1-r^2_p(\mathbf{k},\omega)e^{2ik_{zm}\delta}},\\
\tau_p(\mathbf{k},\omega)&=&(1-r^2_p(\mathbf{k},\omega))\frac{e^{i(k_{zm}-k_z)\delta}}{1-r^2_p(\mathbf{k},\omega)e^{2ik_{zm}\delta}},
\end{eqnarray}
where $r_{\text{TE}}$ and $r_{\text{TM}}$ are the standard vacuum-medium Fresnel reflection coefficients and $k_{zm}=\sqrt{\varepsilon(\omega)\frac{\omega^2}{c^2}-\mathbf{k}^2}$. It is worth stressing at this point that Eqs.~\eqref{alpha1} and \eqref{alpha2} give the total field correlators as a result of four contributions: the blackbody radiation at tempertaure $T_2$, the blackbody radiation reflected by the slab, the blackbody radiation transmitted by the slab and finally the radiation directly emitted by the slab at $T_1$. Note that all but the first contribution depend on the slab properties through the operators $\mathcal{R}$ and $\mathcal{T}$. In particular, in correspondence with a resonance in the dielectric permittivity $\varepsilon(\omega)$ for a value $\omega=\omega_S$ (i.e., in correspondence with a peak in the spectrum of $\varepsilon(\omega)$), the slab-dependent contributions to Eqs.~\eqref{alpha1} and \eqref{alpha2} become dominant for a broad range of atom-slab distances.

Equations \eqref{corr1} and \eqref{corr2} can be used to characterize the influence of atom-field coupling on the atomic dynamics, through the dissipation rates in Eq.~\eqref{MEtot}. Indeed, for real dipoles $\mathbf{d}_i$ of cartesian components $d_i^{x,y,z}$, the rates $\gamma_{ij}^{\pm}(\omega)$ (including $\gamma_i^{\pm}(\omega)=\gamma_{ii}^{\pm}(\omega)$) are \cite{Bellomo2013}
\begin{equation}\label{rates}
\gamma_{ij}^{\pm}(\omega)=\sum_{l,m=x,y,z}c^{ij}_{lm}(\pm \omega)d_i^ld_j^m.
\end{equation}
The rate of absorption and emission of photons from/into the field is the standard way of characterising the field temperature, at least the one perceived by the transition involved in the photons exchange. Introducing the vacuum emission rate $\gamma_0(\omega)=\omega^3(3 \pi \hbar c^3 \varepsilon_0)^{-1}$, one can write
\begin{equation}\label{gammapm}
2\gamma_i^{\pm}(\omega)=\gamma_0(\omega)\big(1\pm1+2n_{\mathrm{env}}(\omega)\big).
\end{equation}
where $n_{\mathrm{env}}(\omega)=\Big[\exp{\big(\hbar \omega/k_B T_{\mathrm{env}}(\omega)\big)}-1\Big]^{-1}$ is the average thermal photon number corresponding to a temperature $T_\text{env}(\omega)$. This allows to characterize the interaction of the OTE field with each atomic transition by means on an effective field temperature. Note however that such temperature fundamentally depends on the transition frequency: two different transitions exchange photons with the same field at different rates and, as such, perceive the same field as having two different temperatures. In particular, thanks to the strong dependence of $\gamma_i^{\pm}$ on the slab dielectric properties, as previously commented, this effective field temperature will be more or less close to the real temperature $T_1$ of the slab, depending on the relative importance of $\alpha_1$ and $\alpha_2$ in Eqs.~\eqref{corr1}-\eqref{corr2}. Therefore, when $\omega=\omega_S$, i.e., the electronic resonance frequency of the slab material, such that both real and imaginary part of its the dielectric permittivity $\varepsilon(\omega)$ show a sharp high peak in their spectrum, the slab contribution to the field correlation functions \eqref{corrfun1}-\eqref{corrfun2} becomes dominant, and the rates \eqref{rates} are profoundly affected by it: transitions at $\omega_S$ feel a temperature much closer to the one of the slab than to the background blackbody radiation.\\
The dipole-dipole coupling strength $\Lambda_{ij}$ has also a similar expression, partly depending on the slab properties and partly originating from a $T=0$ vacuum contribution \cite{Bellomo2013}, which we do not report here for the sake of brevity.
The term $\sum_{i,j}^{\text{res}}\Lambda_{ij}\sigma_i^-\sigma_j^+$ allows two resonant transitions in two different atoms to exchange energy under the form of heat.
\begin{figure}[t!]
\begin{center}
\includegraphics[width=220pt]{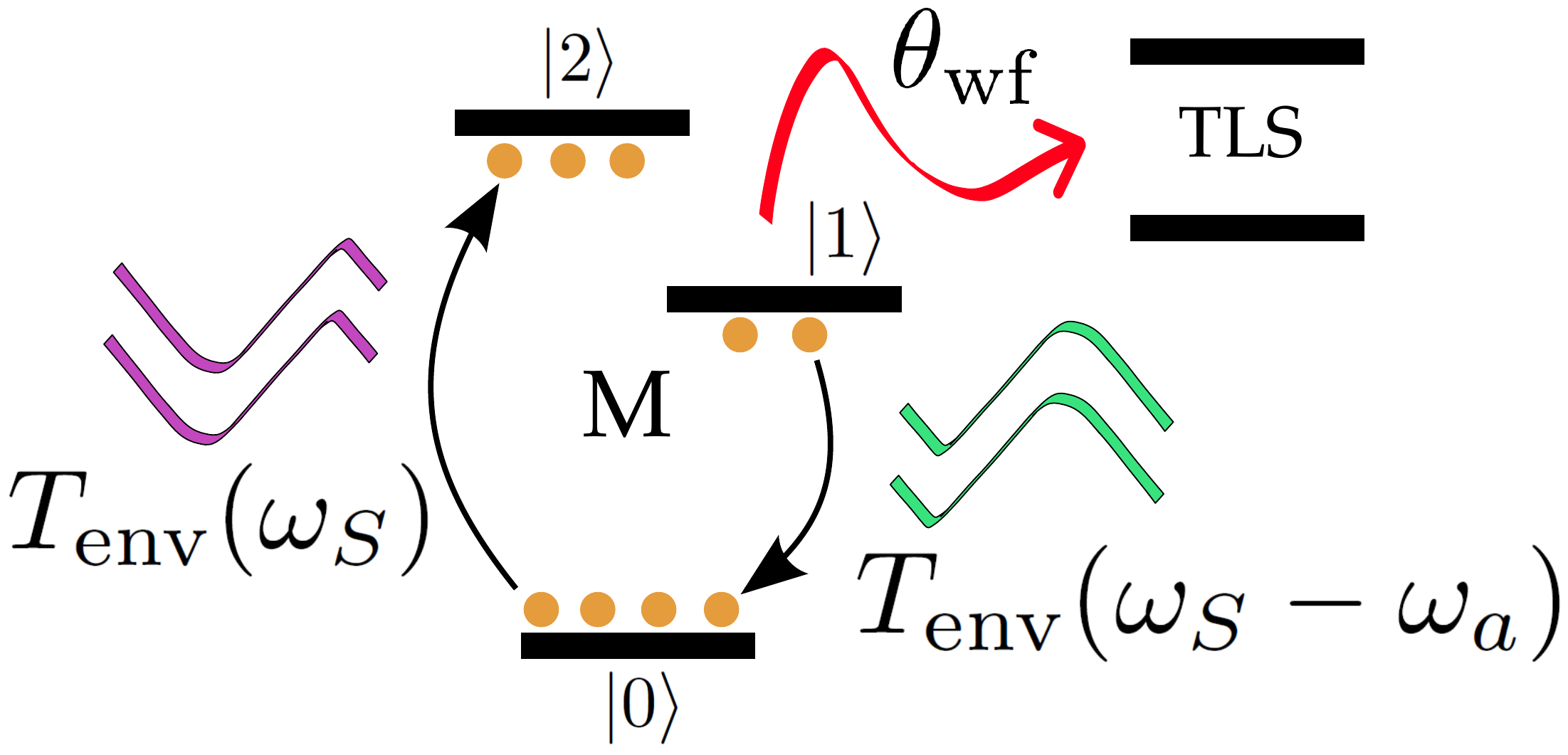}
\end{center}
\caption{Schematic representation of the effects of OTE field in the steady-state of a three level atom M, resonantly coupled to a TLS. The transition $|0\rangle\leftrightarrow|2\rangle$ has the same frequency $\omega_S$ as the electronic resonance of the slab material, while the transition $|1\rangle\leftrightarrow|2\rangle$ is resonant with the TLS at $\omega_a$. Here $T_\z{env}(\omega_S)>T_\z{env}(\omega_S-\omega_s)$ due to a transition-slab resonance effect. In this situation, a redistribution of steady population of M (schematically represented by yellow circles) brings the transition at $\omega_a$ to a much more energetic state, able to induce a steady very high or even negative temperature $\theta_\z{wf}$ to the TLS.}
\label{Figure1OTE}
\end{figure}

Consider now the case of two quantum emitters only, a TLS Q and a three-level system M, placed in this OTE field. M has three non-degenerate transitions 1,2 and 3, one of which (the one at lowest frequency, labeled as 2) is resonant with Q at frequency $\omega_a$. Be now the level structure of M such that the transition 2 connects levels $|1\rangle$ and $|2\rangle$, whereas the high frequency transition be the one between levels $|0\rangle$ and $|2\rangle$, and suppose this latter to be resonant with the slab at $\omega_S$.
Due to the non-trivial dependence of $\gamma_i^{\pm}$ on the transition frequency, the three-level system with three different transitions exchanges photons with the field at different rates and, as such, perceives three different temperatures. In particular, since $T_1>T_2$, the transition $|0\rangle\leftrightarrow|2\rangle$ at $\omega_S$ perceives a much higher effective temperature than the rest of atomic transitions. The situation is therefore somewhat analogous to a three-level system, having each transition connected to a different thermal reservoir.
As explained in \cite{Leggio2015} and schematically shown in Fig.~\ref{Figure1OTE}, the net effect is a redistribution of population in each level of M (represented in Fig.~\ref{Figure1OTE} through yellow circles), leading to a very energetic transition $|1\rangle\leftrightarrow|2\rangle$. As shown in \cite{Leggio2015,Leggio2015b}, due to the fact that this transition of M is resonant with Q, M can deliver into the TLS a large amount of energy through the dipole-dipole interaction $H_\z{eff}$. This energy redistributes the populations in the two energy levels of Q, inducing in it a Gibbs-form steady state $\rho_Q\propto e^{-\hbar \omega_a/k_B \theta_{\text{wf}}}$, corresponding to an atomic temperature $\theta_\z{wf}$ far outside the range $[T_2,T_1]$ and even up to negative values. As such, the net effect of the OTE structure of the field is to allow the temperature of a TLS to be brought to values which would not be accessible just by direct thermal contact of the atom with the real reservoirs at $T_1$ and $T_2$. Note that this is possible only when M and Q are in resonance. Thus, if the frequency of Q were changed to another value $\omega_b$, Q would not interact at all with M and would thus thermalize to the effective temperature $T_{\text{env}}(\omega_b)$.

\section{OTE Quantum Otto Cycle}\label{oteqoc}
We suggest then to exploit this effect to enhance the performances of an Otto cycle using the TLS Q as working fluid. Due to the fundamental role played here by the OTE field, we refer to this modified cycle as OTE quantum Otto cycle.
As commented, this OTE field configuration can be produced by the same two thermal baths considered for the s-QOC, one fixing the temperature of the slab and the other producing the blackbody radiation. The slab is connected to the thermal bath at higher temperature $T_1$, while the one at lower temperature $T_2$ is used to produce a thermal blackbody radiation impinging on the slab itself. To maintain the steady OTE configuration one only needs heat inputs from reservoirs $T_1$ and $T_2$. Such an input will in the following be considered a structural feature of our setup in the form of \textit{housekeeping heat} \cite{Oono1998, Hatano2001}, and thus not taken into account in the evaluation of efficiency, as commonly done in non-equilibrium scenarios \cite{Abah2014}.

\begin{figure}[t!]
\begin{center}
\includegraphics[width=220pt]{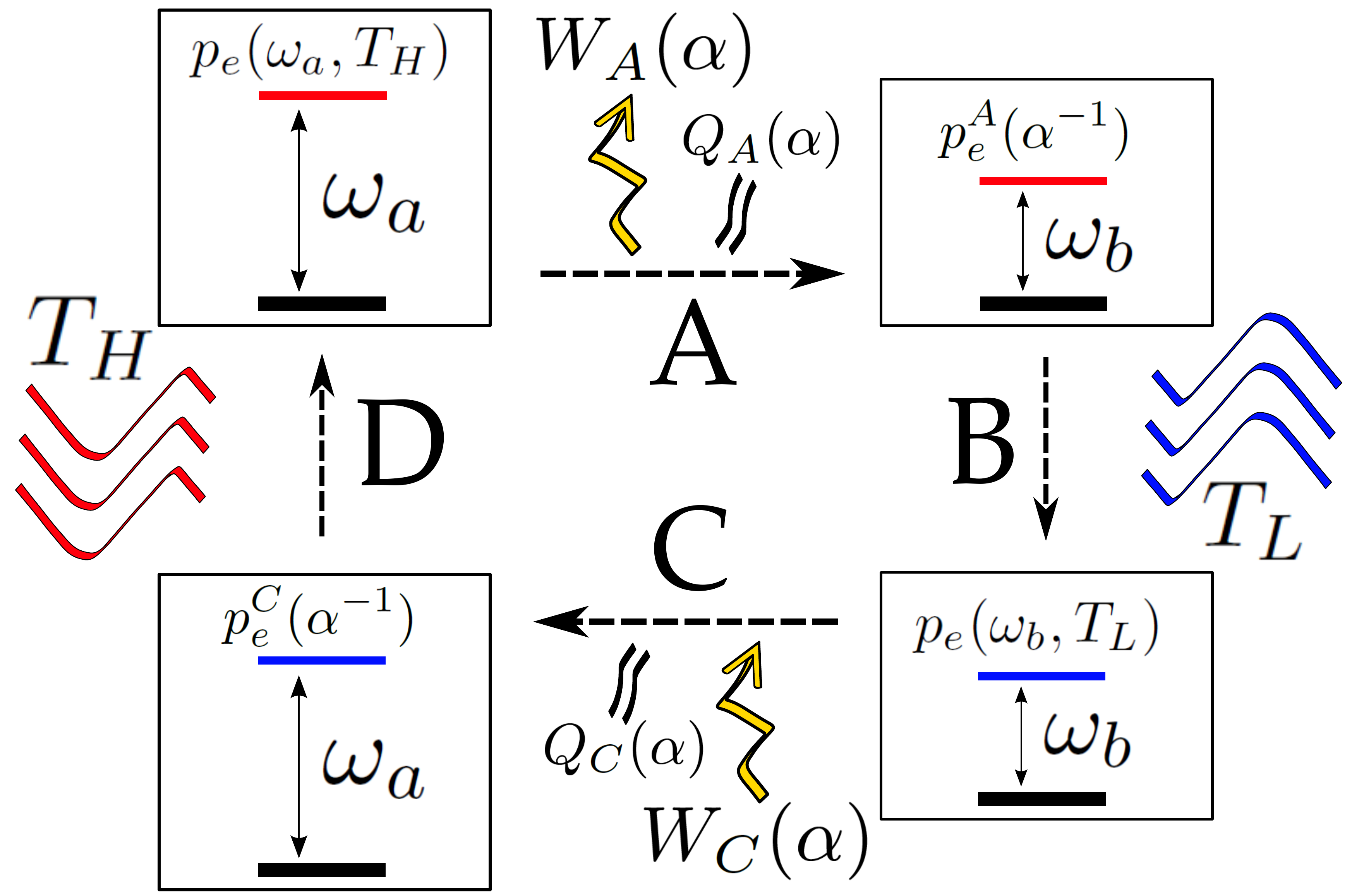}
\end{center}
\caption{Schematic OTE Otto cycle. During stages A and C the two-level working fluid exchanges work with the external world and, possibly, heat with the baths due to the non-perfect adiabaticity of the process. During stages B and D the TLS is put in contact with reservoirs at, respectively, $T_L$ and $T_H$, corresponding respectively to $T_\text{env}(\omega_b)$ and $\theta_\text{wf}$ induced by the OTE field without and with the help of the additional three-level system. The stages A and C are supposed to last for a time $\alpha^{-1}$, such that an ideal cycle is achieved when $\alpha\rightarrow\infty$}
\label{Figure1}
\end{figure}

Due to the OTE properties of the field, as commented in the previous section, the TLS will not interact anymore with reservoirs at $T_1$ and $T_2$ but will rather perceive effective environments depending on its frequency. To stress this difference, we call here $T_H$ and $T_L$ the effective temperatures perceived by the working fluid at, respectively, $\omega_a$ and $\omega_b$, as shown in Fig.~\ref{Figure1}.
Thanks to the presence of M (via atom-atom quantum coherence \cite{Leggio2015}), when the TLS has frequency $\omega_a$ its steady temperature will be $\theta_\text{wf}>T_1$.
Therefore, before stage A begins, the working fluid feels the presence of a much more energetic effective environment than simply the bath at $T_1$, since now $T_H=\theta_{\mathrm{wf}}$. The interaction between M and the TLS is however only possible when the transition of the working fluid is resonant with one of M \cite{Bellomo2013} (incidentally, note that a heat engine based on two resonant emitters in equilibrium baths has been studied in \cite{Klimovsky2015}). Changing $\omega$ from $\omega_a$ to $\omega_b<\omega_a$ puts the TLS and M out of resonance and switches off their interaction. The TLS thus only interacts with the non-equilibrium field, which induces a temperature $T_{\mathrm{env}}\in [T_2,T_1]$ with a non-trivial dependence on the TLS frequency $\omega$. As a consequence, the two new temperatures of the cycle are now $T_H=\theta_{\mathrm{wf}}$ (felt at $\omega_a$) and $T_L=T_{\mathrm{env}}(\omega_b)$.

Consider for instance a slab of SiC ($\omega_S=1.495\times10^{14}\,$rad/s) of $\delta=1\,\mu$m, and be $\omega_a=0.1\times\omega_S$. The three-level atom and the working fluid be at a distance $z=26\,\mu$m from the slab surface and at a distance $r=1\,\mu$m from each other. Solving the long-time limit of Eq.~\eqref{MEtot}, using Eqs.~\eqref{corrfun1}-\eqref{rates} and employing a Drude-Lorentz model for the dielectric permittivity $\varepsilon(\omega)$, one can (numerically) find the two temperatures $T_H$ and $T_L$. When the two external temperatures are $T_1=700\,$K and $T_2=200\,$K, the interaction with M brings the TLS to a temperature $\theta_{\mathrm{wf}}=-537\,$K, i.e., to population inversion. On the other hand, $T_{\mathrm{env}}=313\,$K for $\omega_b=\omega_a/2$.
Note that, strictly speaking, here $T_H<T_L$ since $T_H$ is negative. However, what matters is clearly the fact that the (effective) bath at $T_H$ be more energetic than the one at $T_L$, which is the case here.

With this in mind, let us revisit all the four stages of the Otto cycle in light of this new structure of the (effective) thermal baths of the working fluid:
\begin{itemize}
\item A: $\omega_a\rightarrow \omega_b$ happens now at constant $p_e=p_e(\omega_a,\theta_{\mathrm{wf}})$, higher than the standard value;

\item B: the thermalization changes $p_e(\omega_a,\theta_{\mathrm{wf}})\rightarrow p_e(\omega_b,T_{\mathrm{env}}(\omega_b))$ at constant frequency $\omega_b$;

\item C: $\omega_b\rightarrow \omega_a$ is at constant $p_e=p_e(\omega_b,T_{\mathrm{env}}(\omega_b))$;

\item D: $p_e(\omega_b,T_{\mathrm{env}}(\omega_b))\rightarrow p_e(\omega_a,\theta_{\mathrm{wf}})$ is at constant  $\omega_a$.
\end{itemize}
The work done \textit{by the working fluid} becomes now
\begin{equation}\label{wwf2}
W_{\mathrm{wf}}=\hbar(\omega_b-\omega_a)p_e(\omega_a,\theta_{\mathrm{wf}})+\hbar(\omega_a-\omega_b)p_e(\omega_b,T_L).
\end{equation}
Note that, thanks to the much broader gap between $T_H=\theta_{\mathrm{wf}}$ and $T_L=T_{\mathrm{env}}(\omega_b)$, the PWC in this OTE configuration can become much less restrictive. In particular, when $\theta_{\mathrm{wf}}<0$ and $T_{\mathrm{env}}>0$, work can be extracted from the TLS \emph{for each} value of $\omega_b\leq0$.
Moreover, due to the interaction with M, $p_e(\omega_a,\theta_\text{wf})\gg p_e(\omega_a,T_1)$ such that much more energy is gained when reducing the TLS frequency. On the other hand, $T_{\mathrm{env}}(\omega_b)\in[T_2,T_1]$ by construction, keeping $p_e(\omega_b,T_{\mathrm{env}}(\omega_b))$ relatively closer to $p_e(\omega_b,T_2)$. The net effect is thus to gain an enormous quantity of work compared to the si-QOC case.
The heat absorbed by the TLS is
\begin{equation}
Q_{\mathrm{abs}}=\hbar\omega_a\Big(p_e(\omega_a,\theta_{\mathrm{pw}})-p_e(\omega_b,T_{\mathrm{env}})\Big),
\end{equation}
such that the efficiency is again given by Eq.~\eqref{eta}.
However, when compared to the si-QOC, $\eta$ can now become much higher thanks to the new allowed values for $\omega_b$.

\subsection{Non-ideal OTE Otto cycle} 
Up to now, to preserve adiabaticity we considered an ideal cycle where the TLS frequency is changed suddenly. In realistic models, a finite-time change of frequency corresponds to a non-adiabatic process during which the working fluid exchanges work with the external world and heat with the field reservoir. We account for this by allowing dissipation of the TLS during stages A and C of the cycle. In particular, due to the fact that the dynamics begins when the working fluid is set out-of-resonance with M (stage A) and ends when the two emitters are brought back in resonance (stage C), the dissipative effects are only induced by the electromagnetic field.

We assume a linear time-tuning of the TLS for both stages A and C in the form $\omega_n(t)=\omega_n^{(\mathrm{i})}+(\omega_n^{(\mathrm{f})}-\omega_n^{(\mathrm{i})})\alpha t$, $n=A,C$, and $\omega_{A(C)}^{(\mathrm{i})}=\omega_{a(b)}$, $\omega_{A(C)}^{(\mathrm{f})}=\omega_{b(a)}$. Here $\alpha$ is the adiabatic parameter which characterizes the speed of the stage, in the sense that both stages A and C last for $\alpha^{-1}\,$seconds, and become fully adiabatic when $\alpha\rightarrow\infty$. In the time interval $[0,1/\alpha]$, Eq. \eqref{MEtot} thus reduces to
\begin{equation}\label{ME}
\dot{\rho}_n=\gamma^+\big(\omega_n(t)\big)L(\sigma^-)+\gamma^-\big(\omega_n(t)\big)L(\sigma^+)+U_n(t),
\end{equation}
where $U_n(t)=-i\omega_n(t) \big[\sigma^+\sigma^-,\rho\big]$.

\begin{figure}[h!]
\begin{center}
\includegraphics[width=220pt]{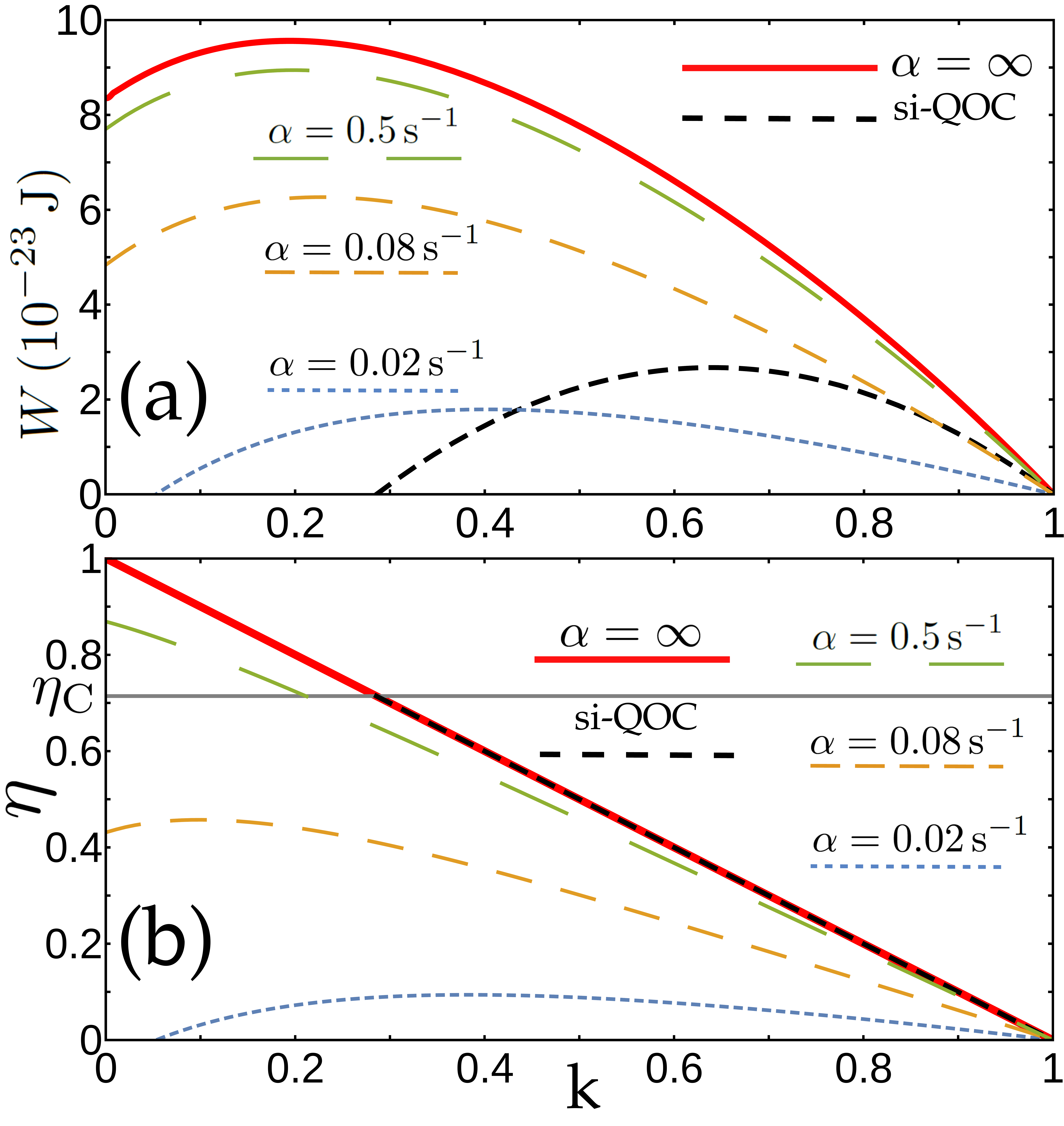}
\end{center}
\caption{Work extracted $W$ [panel (a)] and cycle efficiency $\eta$ [panel (b)] for the OTE Otto cycle versus the ratio $k=\omega_b/\omega_a$. The different curves correspond to different tuning times in stages A and C. Perfect adiabaticity is achieved when $\alpha=\infty$ (full red line). The figure shows also the same quantities for an infinite speed standard quantum Otto cycle (si-QOC, dashed black line) having the same two external temperatures $T_1$ and $T_2$. The standard Carnot efficiency $\eta_C$ is also reported on the left vertical scale and by the horizontal full line in panel (b).\\All the plots are obtained for a SiC slab, with $\delta=1\,\mu$m, $\omega_a=0.1\times\omega_S$, $z=26\,\mu$m, $r=1\,\mu$m, $T_1=700\,$K and $T_2=200\,$K.}
\label{Figure2}
\end{figure}

Solving Eq.~\eqref{ME} with the linear time-dependence of the frequency, one obtains a nontrivial dependence of the excited state population on time. The state of the TLS after stage A or C will thus depend on $\alpha$ and will be referred to as $\rho_{A(C)}(\alpha^{-1})$, with excited state population $p_e^{A(C)}(\alpha^{-1})$, as depicted in Fig.~\ref{Figure1}. Focusing now only on stage A of the cycle (stage C can be treated analogously), the total change in internal energy is $\Delta E_U^{(A)}=\int_0^{1/\alpha}\mathrm{d}t\;\mathrm{tr}\big(\dot{\rho}_AH+\rho_A\dot{H}\big)$. This can be split into a work and a heat part as
\begin{eqnarray}
W_A(\alpha)&=&\int_0^{\frac{1}{\alpha}}\mathrm{d}t\mathrm{tr}\big(\rho_A\dot{H}\big)=\int_0^{\frac{1}{\alpha}}\mathrm{d}t p_e^{(A)}(t)\dot{\omega}_A(t),\\
Q_A(\alpha)&=&\int_0^{\frac{1}{\alpha}}\mathrm{d}t\mathrm{tr}\big(\dot{\rho}_AH\big)=\int_0^{\frac{1}{\alpha}}\mathrm{d}t \dot{p}_e^{(A)}(t)\omega_A(t).
\end{eqnarray}

The total work done by the system during the cycle is $W_{\mathrm{wf}}(\alpha)=W_A(\alpha)+W_C(\alpha)$, whereas the absorbed heat now reads $Q_{\mathrm{abs}}(\alpha)=Q_D+\Theta\big(Q_A(\alpha)\big)Q_A(\alpha)+\Theta\big(Q_C(\alpha)\big)Q_C(\alpha)$, where $\Theta(x)$ is the Heaviside step function of $x$.
Fig.~\ref{Figure2}(a) shows, for an exemplary configuration, the work extracted $-W_{\mathrm{wf}}(\alpha)$ from the TLS at different $\alpha$, together with the same quantity for a si-QOC between the same two temperatures, as a function of the ratio $k=\omega_b/\omega_a$.
For a wide range of values of the adiabatic parameter $\alpha$, the work extracted is much higher than for an \textit{infinite-speed} standard quantum Otto cycle, and is always positive in the whole range $0\leq\omega_b\leq\omega_a$. In particular, the maximum of work extracted in the si-QOC is $W_{\mathrm{max}}^{\mathrm{QOC}}=2.6\times 10^{-23}\,$J, which becomes $W_{\mathrm{max}}^{\mathrm{OTE}}=9.6\times 10^{-23}\,$J for the $\alpha=\infty$ OTE cycle, nearly 4 times bigger. In addition, as shown in Fig.~\ref{Figure2}(b), the OTE efficiency of work extraction asymptotically approaches $1$ as $\omega_b\rightarrow 0$, situation forbidden in the si-QOC due to the value of the PWC.

\begin{figure}[t!]
\begin{center}
\includegraphics[width=220pt]{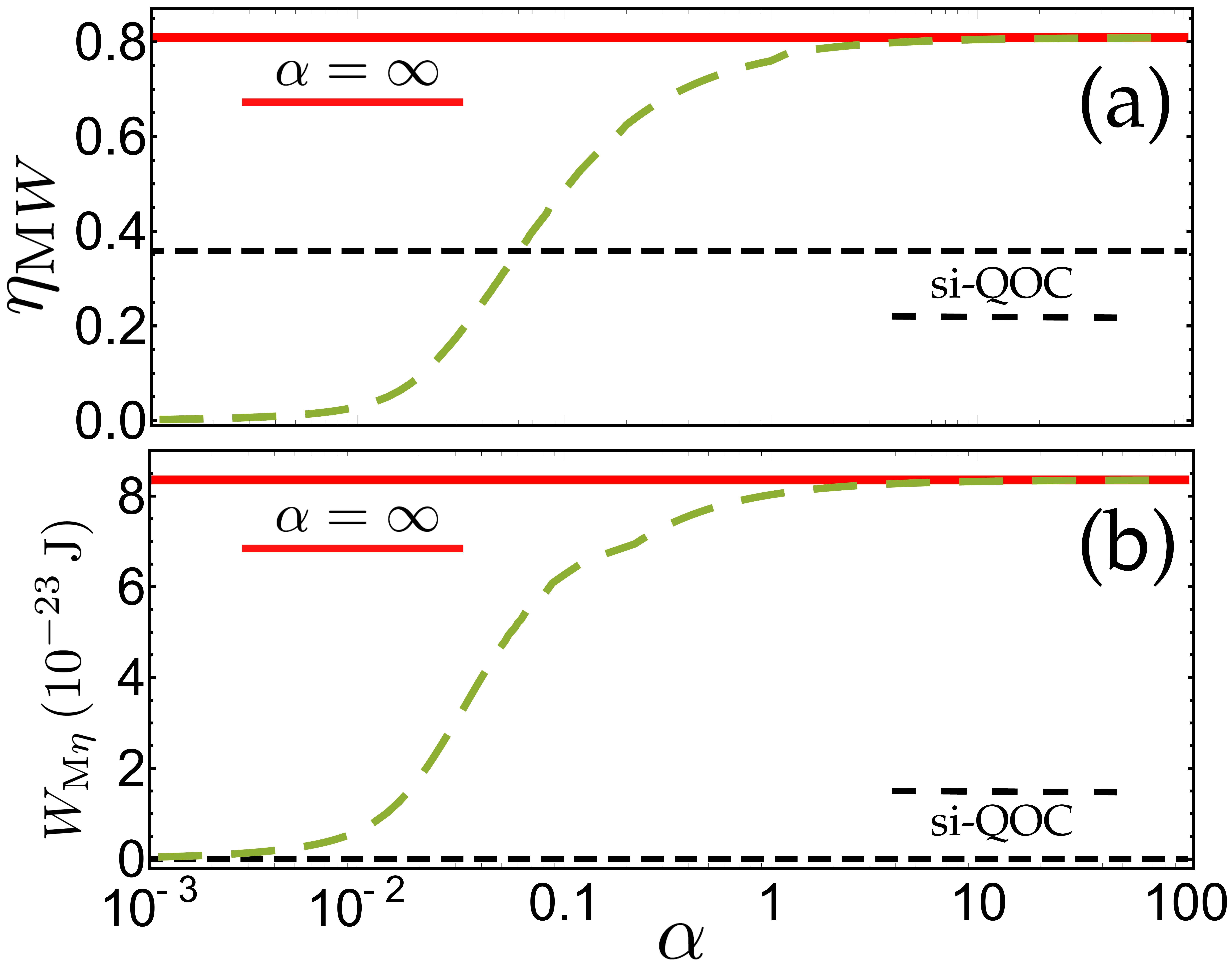}
\end{center}
\caption{Same configuration as in Fig.~\ref{Figure2}: (green dashed line) Efficiency at maximum power $\eta_{\mathrm{M}W}$ (panel (a)) and work at maximum efficiency $W_{\mathrm{M}\eta}$ (panel (b)) for an OTE quantum Otto cycle versus the adiabatic parameter $\alpha$. Both quantities are also shown for a si-QOC between the same two external temperatures. Note that $W_{\mathrm{M}\eta}$ is identically zero for the si-QOC (as expected), but it is always positive for OTE cycles with $\alpha\neq 0$.}
\label{Figure4}
\end{figure}

\begin{figure}[t!]
\begin{center}
\includegraphics[width=220pt]{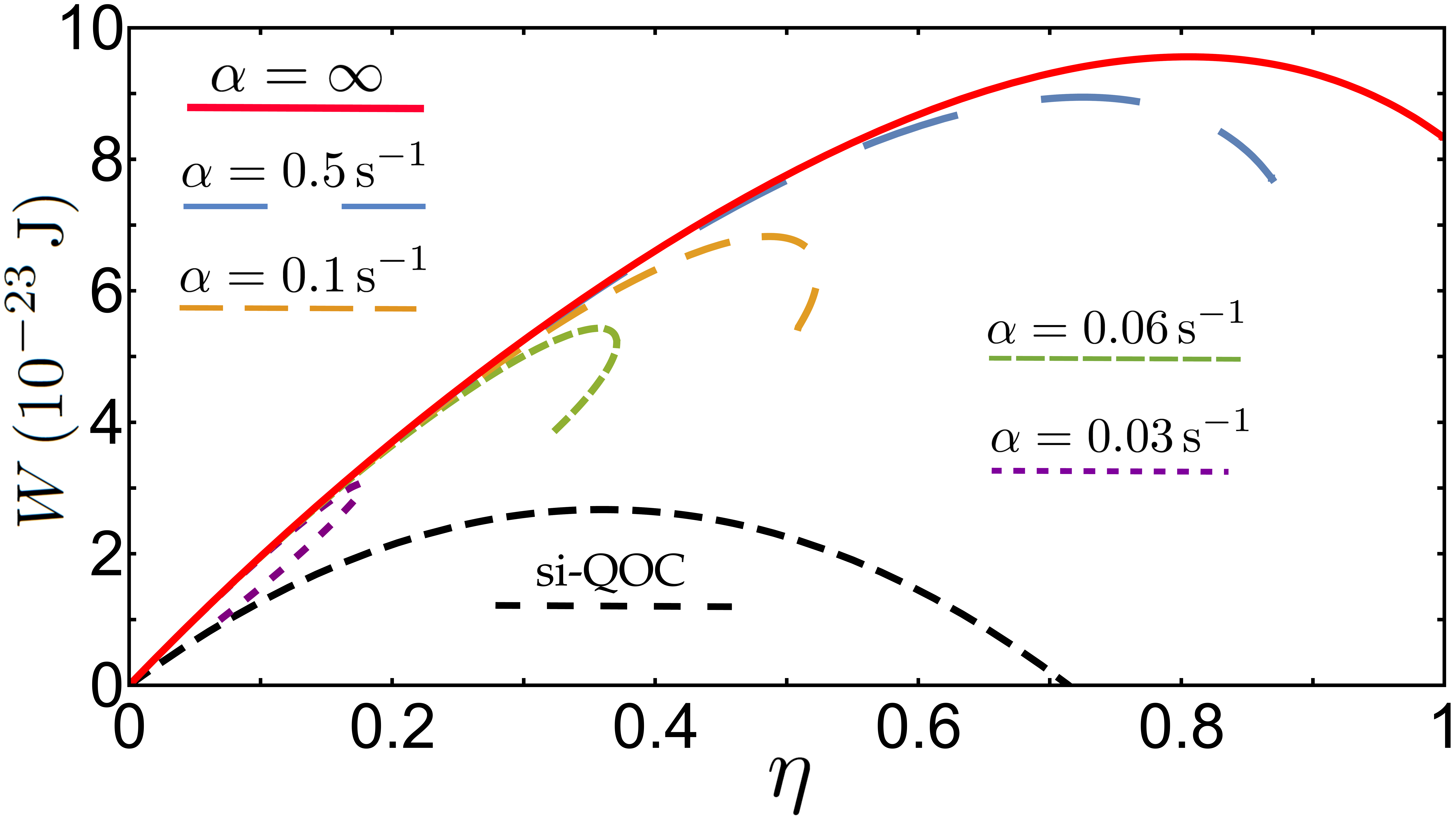}
\end{center}
\caption{Same configuration as in Fig.~\ref{Figure2}: Work extracted versus the corresponding value of efficiency for different values of the cycle adiabatic parameter $\alpha$.}
\label{Figure5}
\end{figure}

Figure~\ref{Figure4} shows, in panel (a) and (b) respectively, the efficiency at maximum power $\eta_{\mathrm{M}W}$ and the work at maximum efficiency $W_{\mathrm{M}\eta}$ for the standard ideal cycle (short-dashed black line), and for both the infinite speed limit (solid red line) and the finite speed (long-dashed green line) OTE cycle. Note that $\eta_{\mathrm{M}W}$ can become greater than its correspondent value for the standard ideal Otto cycle already at finite speed. The infinite-speed limit of the OTE cycle greatly overperforms the infinite-speed standard cycle as $\eta_{\text{M}W}^\text{OTE}>2\eta_{\text{M}W}^\text{QOC}$.
Furthermore, the work at maximal efficiency $W_{\mathrm{M}\eta}$ is of no interest in standard thermodynamic cycles, since it corresponds to the work performed by the cycle working at its Carnot limit, which is known to vanish. In the case of OTE cycles, however, $W_{\mathrm{M}\eta}$ is positive for any non-zero value of $\alpha$, and is very close, for ideal cycles, to $W_{\mathrm{max}}^{\mathrm{OTE}}$. Finite work at asymptotically unitary efficiency is thus the most peculiar characteristics of our OTE cycles, impossible to achieve in standard equilibrium contexts.
The main results are evident in Fig.~\ref{Figure5}, showing for different values of $\alpha$ the curves of $W$ versus the efficiency $\eta$. Remarkably, contrarily to s-QOC, for $\alpha\neq0$ the OTE cycle has non-zero $W_{\mathrm{M}\eta}$, which is a behavior opposite to standard thermodynamic expectations .

\section{Conclusions}\label{conc}
In this work we introduce a quantum Otto cycle scheme which is realized by using a simple non-equilibrium realistic configuration of the electromagnetic field. A two-level system undergoes 4 transformations with the help of a resonant 3-level system. We show that its performances are drastically enhanced, overcoming standard equilibrium thermodynamic bounds. This scheme allows to considerably increase both work extraction and its efficiency. In particular, finite (and almost maximal) work can be extracted at asymptotically unitary efficiency, largely overperforming any standard ideal Otto cycle working between the same two temperatures. The cycle is obtained using a single non-coherent reservoir  produced by heat fluxes provided to macroscopic objects, without the need of any active control on its state.
It exploits quantum atomic coherence which allows the system working between effective thermostats at largely different temperatures. This provides an innovative framework for highly efficient energy management and quantum thermal engines at the quantum scale.

\begin{acknowledgments}
We acknowledge insightful discussions with P. Doyeux and R. Messina, and financial support from the Julian Schwinger Foundation.
\end{acknowledgments}

\end{document}